\documentclass[aps,pra,10pt,reprint,superscriptaddress]{revtex4-2}

\usepackage{natbib}
\usepackage{amsmath,amsfonts,amssymb}
\usepackage{xcolor,graphicx}
\usepackage{tikz}
\usepackage{braket}
\usepackage{dcolumn}
\usepackage{bm}
\usepackage{subfig}
\usepackage{float}
\usepackage[pdftex]{hyperref}
\usepackage{csquotes}
\hypersetup{
	colorlinks = true,
	linkcolor = blue,
	anchorcolor = blue,
	citecolor = red,
	filecolor = blue,
	urlcolor = blue}

\DeclareMathOperator{\ii}{\textrm{i}} 

\begin{document}

\title{Optical ladder operators in the Glauber-Fock oscillator array}

\author{I. Bocanegra}
\email[e-mail: ]{ivanbocanegrag@gmail.com}
\affiliation{Instituto Nacional de Astrofísica Óptica y Electrónica, Calle Luis Enrique Erro No. 1\\ Santa María Tonantzintla, Puebla, 72840, Mexico}
\author{L. Hernández-Sánchez}
\affiliation{Instituto Nacional de Astrofísica Óptica y Electrónica, Calle Luis Enrique Erro No. 1\\ Santa María Tonantzintla, Puebla, 72840, Mexico}
\author{I. Ramos-Prieto}
\affiliation{Instituto Nacional de Astrofísica Óptica y Electrónica, Calle Luis Enrique Erro No. 1\\ Santa María Tonantzintla, Puebla, 72840, Mexico}
\author{F. Soto-Eguibar}
\affiliation{Instituto Nacional de Astrofísica Óptica y Electrónica, Calle Luis Enrique Erro No. 1\\ Santa María Tonantzintla, Puebla, 72840, Mexico}
\author{H. M. Moya-Cessa}
\affiliation{Instituto Nacional de Astrofísica Óptica y Electrónica, Calle Luis Enrique Erro No. 1\\ Santa María Tonantzintla, Puebla, 72840, Mexico}

\date{\today}

\begin{abstract}
In this study, we investigate the stationary states of the Glauber-Fock oscillator waveguide array. We begin by transforming the associated Hamiltonian into the form of a quantum harmonic oscillator Hamiltonian, allowing the implementation of a supersymmetric (SUSY) approach. By considering the simplest case for the intertwining operator, the  optical ladder operators are straightforwardly constructed and shown to map eigensolutions into eigensolutions of the corresponding Hamiltonian operator, in pretty much the same manner as it is done for the quantum harmonic oscillator case. The ladder of the corresponding (eigen) supermodes is then easily established.
\end{abstract}

\maketitle

\section{Introduction}
From the point of view of Supersymmetric (SUSY) quantum mechanics \cite{Witten1981,Andrianov1984,Cooper1995,Bagrov1996,Zelaya2020,CruzyCruz2021}, if there exists an operator $B$, \textit{intertwining} the Hamiltonians $H$ and $\Tilde H$, this is, satisfying 
\begin{equation}\label{inter}
    BH=\Tilde H B,
\end{equation}
then the solutions of the Schrödinger equation associated with $\Tilde H$ can be obtained from those corresponding to $H$ and vice versa: the operator $B$ is then called the intertwiner. Actually, the intertwining relation (\ref{inter}) is often encountered in connection with the so-called \textit{Darboux transformation} \cite{Darboux1882,Bagrov1995,Bagrov1995Russ,Bagrov1997,Fernandez1995,CruzyCruz2020} and the \textit{factorization method} \cite{Schrodinger1940,Schrodinger1940b, Dirac,Infeld1951,Mielnik1984,Mielnik2004,Kuru2008,CruzyCruz2008,CruzyCruz2019,Bocanegra2022,Bocanegra2023}.  Consequently, supersymmetry, the Darboux transformation and the factorization method are sometimes treated as equivalent \cite{Bagrov1995} (and are actually equivalent under certain conditions). 

The relation (\ref{inter}) has been exploited to construct (families of) exactly solvable Hermitian \cite{Mielnik1984,Kuru2008,CruzyCruz2008} and non-Hermitian Hamiltonians \cite{Zelaya2020,CruzyCruz2021,Bocanegra2022,Bocanegra2023}, in the time-independent \cite{Andrianov1984,Bagrov1995,Fernandez1995,CruzyCruz2019} as well as time-dependent cases \cite{Bagrov1995,Bagrov1996,CruzyCruz2020}. In particular, in the stationary regime, if $B = a$ ($B = a^\dagger$), with $a$ ($a^\dagger$) the annihilation (creation) operator of the harmonic oscillator, and $H = n$, with $n=a^\dagger a$ the usual number operator, then $\Tilde H = n+1$ ($\Tilde H = n-1$) \cite{Louisell}. This is, indeed, the basis for the construction of the solution of the eigenvalue equation of the quantum harmonic oscillator developed early by Schrödinger \cite{Schrodinger1940,Schrodinger1940b}, and made popular by Dirac in his book \cite{Dirac}. 

Besides, due to the formal equivalence between the Schrödinger equation and the paraxial Helmholtz equation \cite{Marte1997} (also between the stationary Schrödinger equation and the Helmholtz equation), optical waveguides are suitable devices to observe, study and test quantum phenomena \cite{Longhi2009}. Therefore, either isolated waveguides or waveguide arrays (called optical lattices), are susceptible to be transformed by means of Darboux or SUSY transformations \cite{Miri2013,Miri2014,Heinrich2014,Heinrich2014b, Midya2014,Zuñiga2014,El-Ganainy2015,Contreras-Astorga2019,Queralto2020}, in the Hermitian \cite{Miri2013,Miri2014,Heinrich2014,Heinrich2014b,Zuñiga2014} and non-Hermitian \cite{Midya2014,Contreras-Astorga2019,Queralto2020,Bocanegra2022,Bocanegra2023} regimes. Specifically, an optical lattice associated with a Hamiltonian of the type of the quantum harmonic oscillator can be ``intertwined" with itself, giving rise to the ladder of eigenstates (supermodes \cite{Bocanegra2023b,Bocanegra2023c}).

\begin{figure}[H]
    \centering
    {\includegraphics[width=\linewidth]{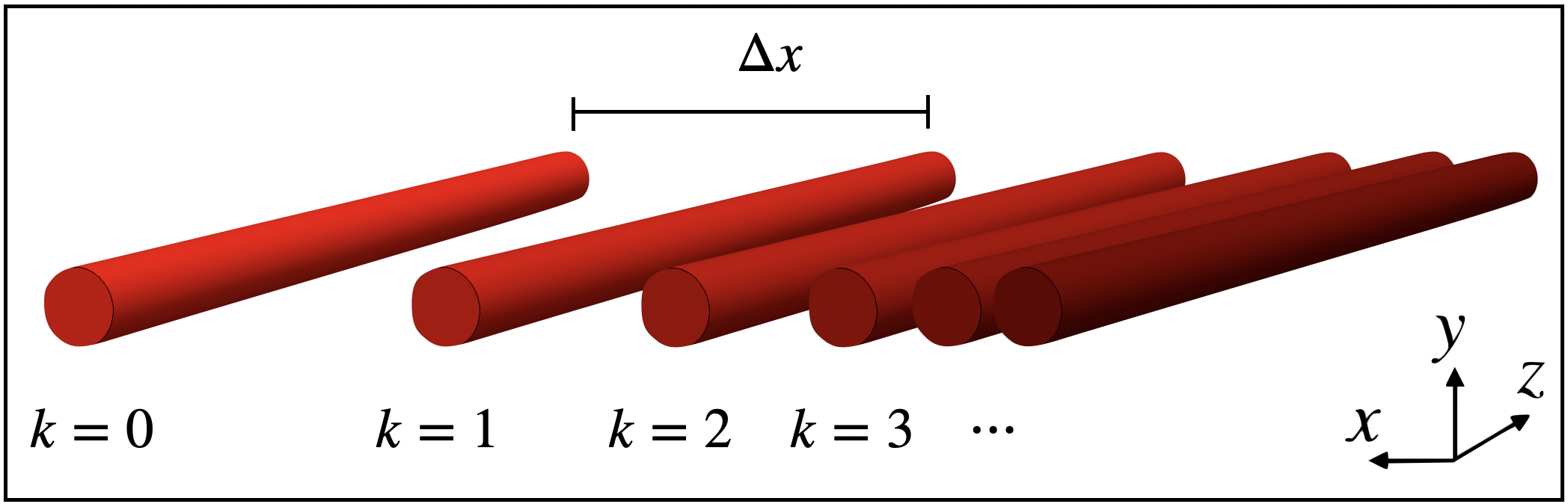}}
\caption{In the Glauber-Fock oscillator array, the waveguides are separated in a non-uniform way, $\Delta x$ is the distance between the first two waveguides. As the site $k$ of the waveguide increases, the waveguides become closer and closer, as shown. In addition, as $k$ grows the refractive index of each waveguide increases as well. Such gradient in refractive index is schematically shown with darker tones of red as $k\to\infty$.}
	\label{fig.guides}
\end{figure}

In the present work we study the stationary regime of a very particular (semi-infinite) waveguide array, referred to as the \textit{Glauber-Fock} (oscillator) \textit{array} \cite{Perez-Leija2012,Keil2012} (also see \cite{Perez-Leija2010,Keil2011,Bocanegra2023b}), and characterized by a non-uniform distance between adjacent waveguides, as well as a gradient of refractive index increasing with the waveguide site $k$, $k=0,1,\dots$ (see Fig. \ref{fig.guides}). The corresponding Hamiltonian $H$ can be taken to the form of the quantum harmonic oscillator Hamiltonian, by means of a rather elementary transformation \cite{Bocanegra2023b}. Then, by considering the simplest case for the intertwiner $B=a$, the operators mapping eigensolutions into eigensolutions of $H$ are straightforward to obtain, and the corresponding ladder of supermodes is easily constructed.

With that in mind, the objective of this paper is twofold. On the one hand, the optical ladder operators associated with $H$ are constructed in a formal way, and interpreted as switchers between the stationary (eigen) supermodes of the Glauber-Fock oscillator array. On the other hand, we set the precedent for further constructions of (Hermitian and non-Hermitian) waveguide arrays that can be obtained by considering more general intertwining operators $B$ (in the stationary as well as the time-dependent regimes). 

The general outline is as follows: in section \ref{sec.introduction} the system under study is introduced, as well as the transformation to turn the corresponding Hamiltonian into the form of the quantum harmonic oscillator Hamiltonian. In section \ref{sec.susy}, the generic SUSY approach is presented, for an arbitrary intertwiner operator $B$. In turn, in section \ref{sec.ladder}, the foundations presented in section \ref{sec.susy} are implemented for the simplest case of the intertwining operator. This results in the optical ladder operators for the stationary eigen supermodes of the Glauber-Fock oscillator array. Finally, in section \ref{sec.conclusiones} the main conclusions are drawn.

\section{Glauber-Fock oscillator array}\label{sec.introduction}
In the Glauber-Fock oscillator array \cite{Perez-Leija2012,Keil2012}, the amplitude of the electric field propagating in the $k$-th waveguide, $k=0,1\dots$, and here denoted $c_k(z)$, is ruled by
\begin{equation}\label{difference}
    \ii \dot{c}_k+ \omega  k c_k +g\left( \sqrt{k}c_{k-1}+\sqrt{k+1}c_{k+1}\right)=0,
\end{equation}
where $\omega\in\mathbb R$ is the propagation constant of the first waveguide,
and $g\in\mathbb R$ is the coupling between the first to waveguides; $g$ is proportional to $e^{-\Delta x}$, with $\Delta x$ the distance between the first two sites (see Fig. \ref{fig.guides}).  Equation (\ref{difference}) can be alternatively written as
\begin{equation}\label{schr}
\ii \partial_z \ket{\psi(z)} =  H \ket{\psi(z)},
\end{equation}
where 
\begin{equation}\label{tdsol}
\ket{\psi(z)} = \sum_{k=0}^\infty c_k(z)\ket{k},
\end{equation}
with $c_j=\braket{ j|\psi(z) }$ for $j=0,1,\dots$. The set $\left\{|k\rangle\right\}_{k=0,1,\dots}$ represents the Fock basis, which spans the Hilbert space $\mathcal{H}$. The Hamiltonian in (\ref{schr}) is given by
\begin{equation}\label{H}
H = -\omega n -g( a^\dagger +  a).
\end{equation}

The state $\ket{\psi(z)}$ in (\ref{tdsol}) contains the information of the total electric field in the array for each $z$. As the Hamiltonian $H$ is time-independent, the mathematical solution of (\ref{schr}), is simply
\begin{equation}\label{sol}
\ket{\psi(z)} = e^{-\ii H z} \ket{\psi(0)}.
\end{equation}
Here we are particularly interested in the initial condition
\begin{equation}
\ket{\psi(0)} = \ket{\psi_\ell },\qquad \ell=0,1,\dots,
\end{equation}
with $\ket{\psi_\ell}$ satisfying the eigenvalue equation
\begin{equation}\label{stat}
H \ket{\psi_\ell} = E_\ell \ket{\psi_\ell},
\end{equation}
and such that the solution (\ref{sol}) is stationary, namely,
\begin{equation}\label{psiz}
    \ket{\psi(z)} = e^{-\ii  H z}\ket{\psi_\ell}= e^{-\ii E_\ell z}\ket{\psi_\ell}.
\end{equation}
By making the transformations
\begin{equation}
    \ket{\psi(z)} = D^\dagger\left(\frac{g}{\omega}\right)\ket{w(z)}, \qquad \ket{\psi_\ell} = D^\dagger\left(\frac{g}{\omega}\right)\ket{ w_\ell},
\end{equation}
where $D(\xi) = \exp(\xi a^\dagger-\xi^* a)$, $\xi\in\mathbb C$, is the Glauber displacement operator \cite{Louisell},  equation (\ref{stat}) and the solution (\ref{psiz}) turn, respectively, into
\begin{equation}\label{w}
    \bar H \ket{w_\ell} = E_\ell \ket{w_\ell},\qquad \ket{w(z)}= e^{-\ii E_\ell z}\ket{w_\ell},
\end{equation}
where
\begin{equation}\label{osc}
    \bar H = D H D^\dagger = -\omega n+\frac{g^2}{\omega},
\end{equation}
is diagonal. The shortcut notation $D=D\left(\frac{g}{\omega}\right)$, $D^\dagger=D^\dagger\left(\frac{g}{\omega}\right)$ is used from now on. The Hamiltonian (\ref{osc}) has the basic form of the quantum harmonic oscillator Hamiltonian, therefore it is susceptible to be supersymmetrically transformed. Next, we set the foundations for a generic SUSY transformation.
\\

\section{Generic SUSY transformation}\label{sec.susy}
If one considers an arbitrary operator $B$ intertwining $\bar H$ in (\ref{osc}) with some other Hamiltonian $\Tilde H$, this is 
\begin{equation}\label{be}
    B\bar H = \Tilde H B,
\end{equation}
the equalities in (\ref{w}) are transformed as
\begin{equation}\label{stat2}
    \Tilde H \ket{\phi_\ell}= E_\ell 
 \ket{\phi_\ell}, \qquad \ket{\phi(z)} = e^{-\ii E_\ell z} \ket{\phi_\ell},
\end{equation}
where we have defined
\begin{equation}\label{phi}
\ket{\phi(z)} =\kappa B \ket{w(z)}, \qquad \ket{\phi_\ell} = \kappa B \ket{ w_\ell},
\end{equation}
with $\kappa\in\mathbb C$ a normalization constant. Therefore, from the first equality in (\ref{osc}), (\ref{be}) becomes the intertwining relation between $H$ and $\Tilde H$:
\begin{equation}
    AH=\Tilde H A,\qquad A=BD,
\end{equation}
and finally, the solution $\ket{\phi_\ell}$ in the second expression of (\ref{phi}) can be written in terms of the solution $\ket{\psi_\ell}$ of the eigenvalue equation (\ref{stat}), as
\begin{equation}
   \ket{\phi_\ell} =\kappa A \ket{\psi_\ell}.
\end{equation}
In turn, the stationary evolution is given by the second expression in (\ref{stat2}).

Therefore, in order to perform the supersymmetric transformation, the solutions of the eigenvalue equation (\ref{stat}) must be known. The obtention of both the set of eigenvalues and eigenvectors of (\ref{stat}), $E_\ell$ and $\ket{\psi_\ell}$, respectively, are given in the Appendix \ref{append}.

In what follows, the simplest case for the intertwiner $B$ is considered, the Hamiltonian $H$ is ``intertwined" with itself, giving rise to the straightforward construction of optical ladder operators connecting solutions of (\ref{stat}), as well as the corresponding ladder of eigen supermodes.

\section{Construction of the ladder operators}\label{sec.ladder}
\begin{figure}[H]
    \centering
    {\includegraphics[width=\linewidth]{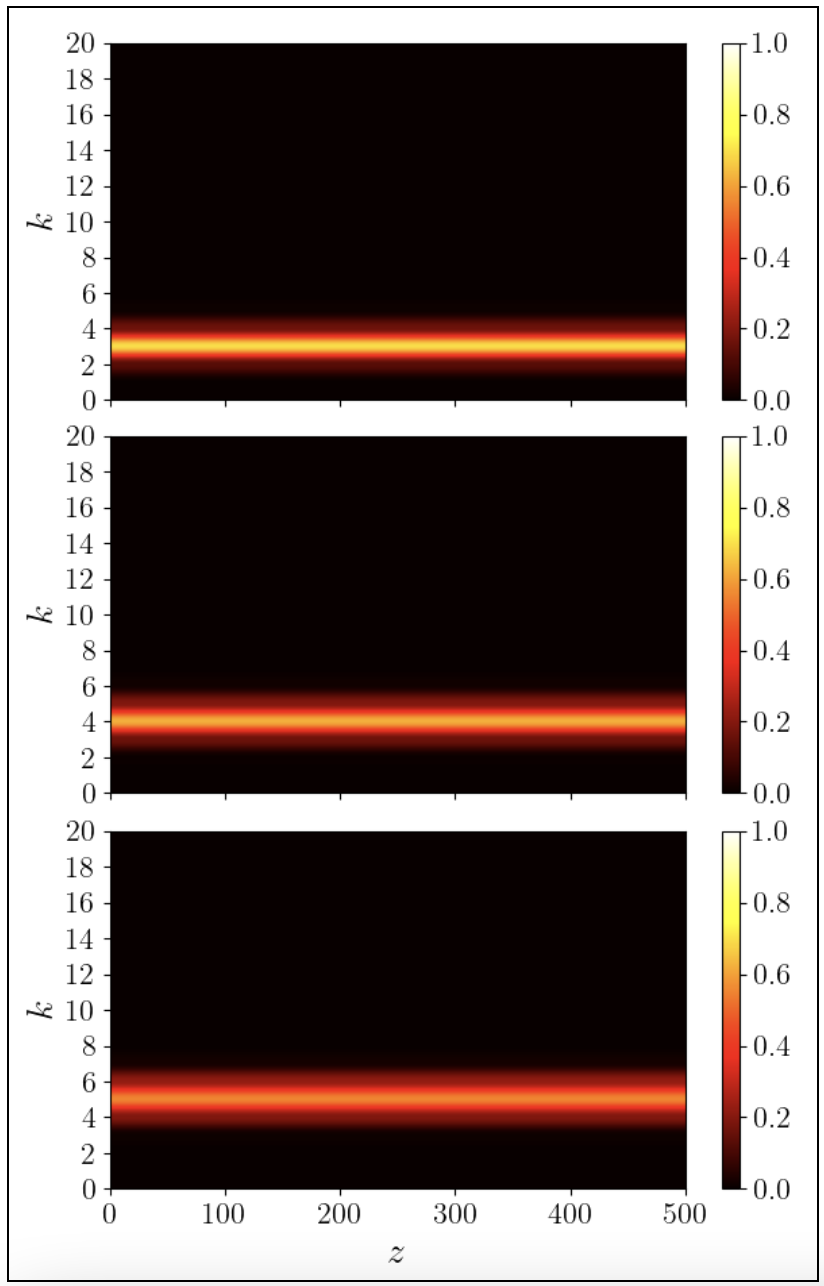}}
\caption{Effect of the optical ladder operators (\ref{km1}) and (\ref{kp1}) on a given initial eigen supermode of the Glauber-Fock oscillator array. The chosen parameters are: $\omega=1.3$, $g=0.29$. By departing from the normalized eigen supermode $\ket{\psi_4}$ (middle), the (also normalized) state $\ket{\psi_3}$ (up) can be obtained by means of the optical annihilation operator (\ref{km1}), while the normalized state $\ket{\psi_5}$ (down) is obtained through the optical creation operator (\ref{kp1}).}
	\label{fig.prop}
\end{figure}

In this section, the case $B=a$ ($B=a^\dagger$) is studied, as we are interested in the construction of the ladder of (eigen) supermodes associated with the Hamiltonian (\ref{H}). Nevertheless, the present discussion is intended to set the precedent for more general choices of the intertwiner $B$, as explained in Section \ref{sec.susy}. From (\ref{osc}), it is easy to prove that 
\begin{equation}\label{as}
    a\bar H = (\bar H-\omega)a,\qquad a^\dagger \bar H = (\bar H +\omega)a^\dagger.
\end{equation}
From the point of view of SUSY quantum mechanics [compare (\ref{be}) with both expressions in (\ref{as})], this means that $a$ ($a^\dagger$) intertwines $\bar H$ with itself (see Ref. \cite{Infeld1951}). From both expressions in (\ref{as}), and by using the first equality in (\ref{osc}), we obtain, respectively
\begin{equation}
    (D^\dagger a D) H = (H-\omega)(D^\dagger a D)
\end{equation}
and
\begin{equation}
    (D^\dagger a^\dagger D) H = (H+\omega)(D^\dagger a^\dagger D),
\end{equation}
where $D^\dagger a D = a + \frac{g}{\omega}$ and $D^\dagger a^\dagger D = a^\dagger + \frac{g}{\omega}$ are the optical ladder operators connecting the stationary (eigen) supermodes of $H$. By applying the operator $D^\dagger a D$ to the left of equation (\ref{stat}), it is obtained
\begin{equation}
    H \left[D^\dagger aD \ket{\psi_\ell}\right] = (E_\ell+\omega)\left[D^\dagger aD \ket{\psi_\ell}\right],
\end{equation}
where
\begin{equation}\label{km1}
    (D^\dagger a D) \ket{ \psi_\ell} = e^{\ii \omega z} \sqrt{\ell} \ket{\psi_{\ell-1}}.
\end{equation}
Similarly, by acting with $D^\dagger a^\dagger D$ on the left of (\ref{stat}), we obtain
\begin{equation}
    H \left[D^\dagger a^\dagger D \ket{\psi_\ell}\right] = (E_\ell-\omega)\left[D^\dagger a^\dagger D \ket{\psi_\ell}\right],
\end{equation}
where
\begin{equation}\label{kp1}
    (D^\dagger a^\dagger D) \ket{\psi_\ell} = e^{-\ii \omega z} \sqrt{\ell+1} \ket{\psi_{\ell+1}}.
\end{equation}
From expressions (\ref{km1}) and (\ref{kp1}) it can be seen that the operators $D^\dagger a^\dagger D$ and $D^\dagger a^\dagger D$ are indeed ladder operators, switching between eigensolutions of (\ref{stat}), and therefore give the stationary evolution dictated in (\ref{psiz}). The ladder of supermodes can be straightforwardly constructed. Similar to the quantum harmonic oscillator case, the $\ell$-th eigenstate (supermode) of the ladder, can be written in terms of the lowest state $\ket{\psi_0}$, as
\begin{equation}
    \ket{\psi_\ell} = \frac{e^{\ii \omega \ell z}}{\sqrt{\ell!}}D^\dagger (a^\dagger)^\ell D \ket{\psi_0}.
\end{equation}
Figure \ref{fig.prop} shows the effect of the optical annihilation and creation operators (\ref{km1}) and (\ref{kp1}), respectively, for some specific values of the parameters. By departing from the normalized eigen supermode $\ket{\psi_4}$ in Fig. \ref{fig.prop} (middle), the (normalized) state $\ket{\psi_3}$ in Fig. \ref{fig.prop} (up) is obtained by means of the application of the operator (\ref{km1}). In turn, again departing form $\ket{\psi_4}$, the normalized eigen supermode $\ket{\psi_5}$ in Fig. \ref{fig.prop} (down) is reached through the operator (\ref{kp1}).

\begin{figure}[H]
    \centering
    {\includegraphics[width=\linewidth]{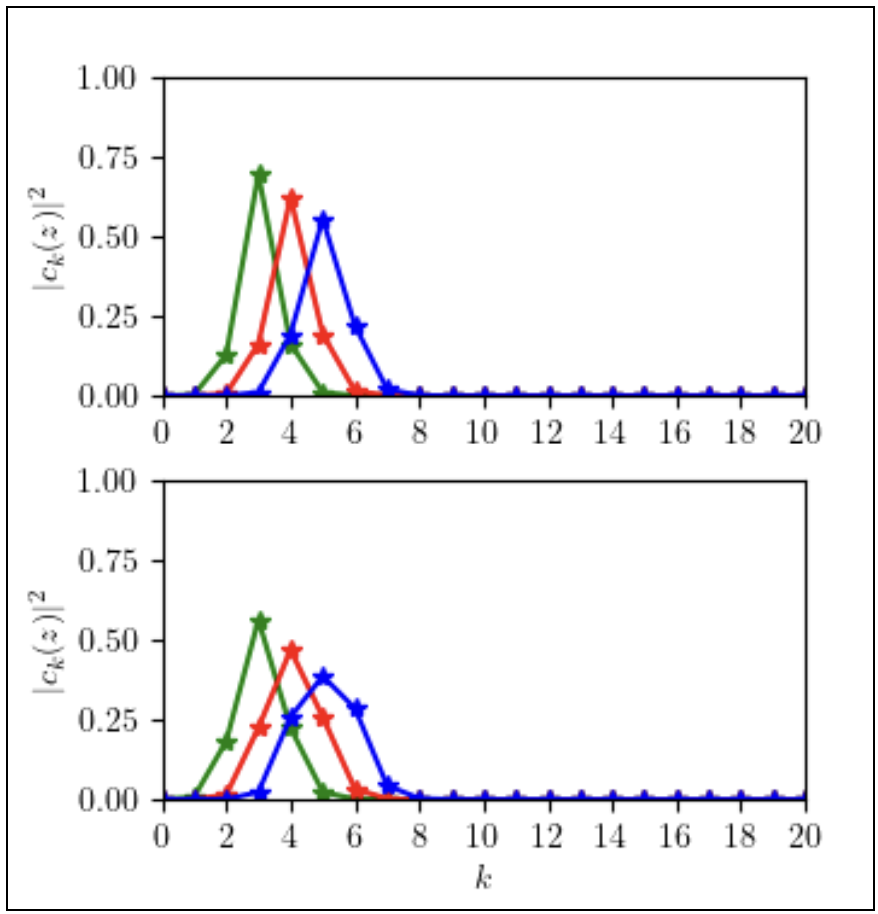}}
\caption{Up: Distributions $c_k$ corresponding to the eigen supermodes $\ket{\psi_3}$ (green), $\ket{\psi_4}$ (red) and $\ket{\psi_5}$ (blue) in Fig. \ref{fig.prop} (up), Fig. \ref{fig.prop} (middle) and Fig. \ref{fig.prop} (down), respectively. As the reader can appreciate, as $\ell$ increases ($\ell=3,4,5$), the distribution becomes smaller and broader. This is in agreement with Fig. \ref{fig.prop}. Down: Distributions $c_k$ for $g=0.36$. As can be seen, as the coupling $g$ grows, the distributions become broader as well. This can be particularly observed in the eigen supermode $\ket{\psi_5}$ (blue). The solid lines correspond to the analytical results while the the stars correspond to the numerical solutions.}
	\label{fig.in}
\end{figure}

In addition, Fig. \ref{fig.in} (up) shows the normalized distributions $c_k$ corresponding to the three eigenstates ($\ell=3,4,5$) shown in Fig. \ref{fig.prop}. The green, red and blue lines correspond, respectively, to $\ket{\psi_3}$ [Fig. \ref{fig.prop} (up)],
$|\psi_4\rangle$ [Fig. \ref{fig.prop} (middle)]
 and $\ket{\psi_5}$ [Fig. \ref{fig.prop} (down)]. It can be observed that, as $\ell$ increases, the distribution $c_k$ becomes smaller and broader, in agreement with Fig. \ref{fig.prop}. In turn, Fig. \ref{fig.in} (down) shows $c_k$ for a slightly greater coupling $g=0.36$. It is seen that, as the coupling $g$ grows, the distributions become broader as well. This can be particularly appreciated in the $\ket{\psi_5}$ supermode (blue) in Fig. \ref{fig.in} (down). In both cases the solid lines correspond to the analytical results while the the stars correspond to the numerical solutions. 

\section{Conclusions}\label{sec.conclusiones}
The optical ladder operators for the Glauber-Fock oscillator waveguide array were constructed in the stationary regime. They switch between eigen supermodes, as expected. In particular, when $g$ is relatively smaller, they switch between the excited waveguide, as shown in Figure \ref{fig.prop}. Therefore, the ladder of eigen supermodes is straightforwardly constructed by departing from the lowest state, in much the same manner as in quantum harmonic oscillator case. 

In addition, as the ladder operators were constructed by departing from a supersymmetric (SUSY) approach, a different choice of the intertwiner might lead to new configurations of waveguide arrays. These might possibly be associated with non-Hermitian or even with time-dependent Hamiltonians. Work in these direction is already in progress.
\\
\appendix
\section{Solution of the initial eigenvalue equation}\label{append}
By using Eq. (3.1.12) of Ref. \cite{Louisell}, from (\ref{sol}) and (\ref{osc}), it is straightforward to obtain
\begin{equation}\label{driv}
\ket{\psi(z)}=D^\dagger\left(\frac{g}{\omega}\right)e^{\ii z(\omega n-\frac{g^2}{\omega})}D\left(\frac{g}{\omega}\right)\ket{\psi(0)}.
\end{equation}
By inspection it is seen that, by choosing $\ket{\psi(0)} = D^\dagger\left(\frac{g}{\omega}\right)\ket{\ell}$, with $\ket{\ell}$ a Fock state satisfying
\begin{equation}
    n\ket{\ell}=\ell\ket{\ell},
\end{equation}
the stationary solution $\ket{ \psi(z)} =\ket{ \psi_\ell}$, with
\begin{equation}\label{eigenvec}
    \ket{ \psi_\ell} = e^{\ii z(\omega n-\frac{g^2}{\omega})} D^\dagger\left(\frac{g}{\omega}\right)\ket{ \ell},
\end{equation}
is obtained. In turn, by acting with $H$ on the state (\ref{eigenvec}), the corresponding spectrum 
\begin{equation}\label{eigenval}
    E_\ell = -\omega \ell +\frac{g^2}{\omega},
\end{equation}
is reached.

\section*{Acknowledgments}
I. Bocanegra acknowledges CONAHCyT (M\'exico) for financial support through the postdoctoral fellowship 711878 and projects A1-S-24569 and CF19-304307. He is also grateful to IPN (M\'exico) for supplementary economical support through the project SIP20232237. L. Hernández Sánchez also thanks the Instituto Nacional de Astrofísica, Óptica y Electrónica (INAOE) and the Consejo Nacional de Humanidades, Ciencias y Tecnologías (CONAHCyT) for the PhD scholarship awarded (No. CVU: 736710).


\end{document}